\documentclass[12pt]{article}
\begin{document}
%%%%%%%%%%%%%%%%%%%%%%%%%%%%%%%%%%%%%%%%%%%%%
\baselineskip 15pt
%%%%%%%%%%%%%%%%%%%%%%%%%%%%%%%%%%%%%%%%%%%%%
\centerline{\Large{\bf On recurrence equations associated with}}
\centerline{\Large{\bf invariant varieties of periodic points}}

%%%%%%%%%%%%%%%%%%%%%%%%%%%%%%%%%%%%%%%%%%%%%
\vglue1cm
\centerline{Satoru SAITO~$^\dag$ and Noriko SAITOH~$^\ddag$}

\centerline{$^\dag$~Hakusan 4-19-10, Midori-ku, Yokohama 226-0006, Japan} 

\centerline{email: saito@phys.metro-u.ac.jp} 

\centerline{$^\ddag$~Applied Mathematics, Yokohama National University,}

\centerline{Hodogaya-ku, Yokohama 240-8501, Japan}

\centerline{email: nsaitoh@ynu.ac.jp}
\vglue1cm
\begin{center}
\begin{minipage}{14cm}
\noindent
{\bf Abstract}
A recurrence equation is a discrete integrable equation whose solutions are all periodic and the period is fixed. We show that infinitely many recurrence equations can be derived from the information about invariant varieties of periodic points of higher dimensional integrable maps. 
\end{minipage}
\end{center}

%%%%%%%%%%%%%%%%%%%%%%%%%%%%%%%%%%%%%%%%%%%%%%%
\section{Introduction}

A recurrence equation is a discrete integrable equation whose solutions are all periodic and the period is fixed. Some of them had been known for some years, while some others have been found recently. In this contribution we would like to show that infinitely many recurrence equations can be derived from the information about invariant varieties of periodic points of higher dimensional integrable maps. Especially recurrence equations associated with the Quispel, Roberts and Thompson (QRT) map\cite{QRT} are shown to exist one for each period.

Some examples of the recurrence equations are \cite{GKP}
\begin{eqnarray}
x_{n+1}&=&\frac {a}{x_n},\qquad a:\ {\rm constant}\label{2}\\
x_{n+1}&=&{1+x_n\over x_{n-1}},\label{5}\\
x_{n+1}&=&{1+x_n+x_{n-1}\over x_{n-2}}.\label{8}
\end{eqnarray}
An interesting feature of these equations is that, for an arbitrary initial value, the solution is always periodic with a fixed period. The period is 2 in the case of (\ref{2}), 5 in the case of (\ref{5}) and 8 in the case of (\ref{8}). These equations were named the recurrence equations by the authors of \cite{HY,HT} who found many other examples of this type recently.

Apparently the recurrence equations are integrable. However there has not been known, to our knowledge, any method to find them systematically. The purpose of this article is to develop a method to derive the recurrence equations from integrable maps in a systematic way.\\

Our key observation is that, writing $(x_{n},x_{n-1},x_{n-2})$ as $(x,y,z)$, the recurrence equations (\ref{2}), (\ref{5}), (\ref{8}) are equivalent to the higher dimensional maps
\begin{eqnarray}
x&\rightarrow& X={a\over x}
\label{2a}\\
(x,y)&\rightarrow&(X,Y)=\left({1+x\over y},\ x\right)
\label{5a}\\
(x,y,z)&\rightarrow& (X,Y,Z)=\left({1+x+y\over z},\ x,\ y\right),
\label{8a}
\end{eqnarray}
respectively. There are a pair of fixed points at $x=\pm\sqrt a$, $x=y=(1\pm\sqrt 5)/ 2$, and $x=y=z=1\pm\sqrt 2$ for each map (\ref{2a}), (\ref{5a}), (\ref{8a}). Otherwise an arbitrary point on the complex space can be an initial point of the periodic map of the corresponding period.\\

We have shown, in our recent paper\cite{SS,SS2}, that periodic points of higher dimensional integrable maps with some invariants form an invariant variety for each period. The invariant variety of periodic points is determined by imposing certain relations among the invariants. Every point on an invariant variety can be an initial point of the periodic map of the same period. All images of the map stay on this invariant variety. Therefore the map defines a recurrence equation of the fixed period if it is constrained on the invariant variety. In some cases the invariant varieties can be derived iteratively for all periods. We can associate one recurrence equation to every invariant variety, thus obtain infinitely many recurrence equations.

We explain briefly the notion of invariant varieties of periodic points in \S 2. Many recurrence equations associated with the invariant varieties will be derived in \S 3. We discuss, in \S 4, a method which enables us to derive series of recurrence equations.

%%%%%%%%%%%%%%%%%%%%%%%%%%%%%%%%%%%%%%%%%%%%%%%%%%%%
\section{Invariant varieties of periodic points}

Let us consider an iteration of a rational map on $\mathbf{\hat C}^d$, where $\mathbf{\hat C}=\{\mathbf{C},\infty\}$,
\begin{equation}
\mathbf{x}=(x_1,x_2,...,x_d)\quad \rightarrow\quad
\mathbf{X}=(X_1,X_2,...,X_d)=:\mathbf{X}^{(1)},
\label{x->X}
\end{equation}
and assume $H_1({\mathbf x}),H_2({\mathbf x}),...,H_p({\mathbf x})$ be the $p$ invariants. We are interested in the behaviour of periodic points satisfying the conditions
\begin{equation}
{\mathbf X}^{(n)}={\mathbf x},\qquad n=2,3,....
\label{X^n=x}
\end{equation}

If $h_1,h_2,\cdots,h_p$ are the values of the invariants determined by the initial point, the map (\ref{x->X}) is constrained on the $d-p$ dimensional algebraic variety $V(h)$, 
\begin{equation}
V(h)=\Big\{\mathbf{x}\Big|\ H_i(\mathbf{x})=h_i,\ i=1,2,...,p\Big\},
\label{V(h)}
\end{equation}
and the periodicity conditions (\ref{X^n=x}) are reduced to the constraints on some $d-p$ functions $\Gamma^{(n)}_\alpha$:
\begin{equation}
\Gamma^{(n)}_\alpha(h_1,h_2,...,h_p,\xi_1,\xi_2,...,\xi_{d-p})=0,\qquad \alpha=1,2,...,d-p,\quad n\ge 2.
\label{Gamma_n=0}
\end{equation}
Here by $\xi_1,\xi_2,...,\xi_{d-p}$ we denote the variables which parameterize the variety $V(h)$ after the elimination of the $p$ components of $\mathbf{x}$. Note that the fixed point conditions $(n=1)$ are excluded in (\ref{Gamma_n=0}) since they are nothing to do with the invariants.

For an arbitrary set of values of $h_1,h_2,...,h_p$, the functions $\Gamma_n^{(\alpha)}(h,\mathbf{\xi})$ define an affine variety, which we denote by $V^{(n)}(\langle \Gamma\rangle)$, {\it i.e.},
\begin{eqnarray*}
V^{(n)}(\langle\Gamma\rangle)=\Big\{\mathbf{\xi}\Big|\ \Gamma^{(n)}_\alpha(h,\mathbf{\xi})=0,\ \ \alpha=1,2,...,d-p\Big\},\quad n\ge 2.
\end{eqnarray*}
In general this variety consists of a finite number of isolated points on $V(h)$, hence zero dimension, corresponding to the solutions to the $d-p$ algebraic equations (\ref{Gamma_n=0}) for the $d-p$ variables $\xi_1,\xi_2,...,\xi_{d-p}$. In this case we say that the periodicity conditions (\ref{X^n=x}) are `uncorrelated'. If the values of the invariants are changed continuously these points move all together and form a subvariety of dimension $p$ in $\mathbf{\hat C}^d$. Needless to say this case includes a map with no invariant. \\

There are possibilities that the equations $(\ref{Gamma_n=0})$ impose relations on $h_1,h_2,...,h_p$ instead of fixing all $\xi_\alpha$'s. Let $l$ be the number of such equations. We write them as
\begin{equation}
\gamma^{(n)}_\alpha(h_1,h_2,...,h_p)=0,\qquad \alpha=1,2,...,l,
\label{gamma_n}
\end{equation}
instead of $\Gamma^{(n)}_\alpha$, to emphasize independence from $\xi_j$'s. If $m$ is the number of the rest of the equations
\begin{equation}
\Gamma^{(n)}_\alpha(h_1,h_2,...,h_p,\xi_1,\xi_2,...,\xi_{d-p})=0,\qquad \alpha=1,2,...,m
\label{Gamma_n}
\end{equation}
$d-p-m$ variables are not determined from the periodicity conditions. This means that $V^{(n)}(\langle\Gamma\rangle)$ forms a subvariety of dimension $d-p-m$ of $V(h)$.
We say that the periodicity conditions are `correlated' in this case. In \cite{SS} we have proved the following lemma:\\

\noindent
{\bf Lemma}\cite{SS}

{\it A set of correlated periodicity conditions satisfying $
\min\{ p,d-p\} \ge l+m
$
and a set of uncorrelated periodicity conditions of a different period do not exist in one map simultaneously.}\\

When $m=0$, in particular, the periodicity conditions determine none of the variables $\xi_1,\xi_2,...,\xi_{d-p}$ but impose $l$ relations among the invariants. Then the affine variety $V^{(n)}(\langle\Gamma\rangle)$ coincides with $V(h)$. In other words every point on $V(h)$ is a periodic point of period $n$, while $V(h)$ itself is constrained by the relations among the invariants. We say the periodicity conditions are `fully correlated' in this particular case. If we replace $h_i$ by $H_i({\mathbf x})$ in $\gamma^{(n)}_\alpha(h)$ the periodicity conditions (\ref{gamma_n}) enable us to consider the constraints on the invariants as constraints on the variables $\mathbf{x}$. We denote by $v^{(n)}(\langle \gamma\rangle)$ the affine variety generated by the functions $\gamma^{(n)}_\alpha(H_1(\mathbf{x}),H_2(\mathbf{x}),...,H_p(\mathbf{x}))$, and distinguish it from $V^{(n)}(\langle \Gamma\rangle)$. Namely we define
\begin{eqnarray}
v^{(n)}(\langle\gamma\rangle)=\Big\{\mathbf{x}\Big|\ \gamma^{(n)}_\alpha(H_1(\mathbf{x}),H_2(\mathbf{x}),...,H_p(\mathbf{x}))=0,\ \ \alpha=1,2,...,l\Big\}.
\label{ivpp}
\end{eqnarray}

We call $v^{(n)}(\langle \gamma\rangle)$ `an invariant variety of periodic points', whose properties can be summarized as follows:
\begin{itemize}
\item
The dimension of $v^{(n)}(\langle \gamma\rangle)$ is $d-l\ (\ge p)$.
\item
Every point on $v^{(n)}(\langle \gamma\rangle)$ can be an initial point of the periodic map of period $n$.
\item
All images of the periodic map starting from a point of $v^{(n)}(\langle \gamma\rangle)$ stay on it.
\item
$v^{(n)}(\langle \gamma\rangle)$ is determined by the invariants of the map alone.
\end{itemize}
\vglue0.5cm

If the periodicity conditions of period $n$ are fully correlated, {\it i.e.}, when $m=0$ in (\ref{Gamma_n}), the condition $\min\{ p,d-p\} \ge l+m$ is always satisfied as long as (\ref{gamma_n}) has solutions. Our theorem thus follows from the lemma and this fact immediately.\\

\noindent
{\bf Theorem}\cite{SS}

{\it 
If there is an invariant variety of periodic points of some period, there is no set of isolated periodic points of other period in the map.
}\\

This theorem tells us nothing about the integrability of a map. To proceed further we assume that a nonintegrable map has at least one set of uncorrelated periodicity conditions. This is certainly true if the map has a Julia set. Once we adopt this observation as a working hypothesis, our theorem is equivalent to the following statement:\\

\noindent
{\it If a map has an invariant variety of periodic points of some period, it is integrable.}\\

In order to support this proposition we have investigated various known integrable maps and found invariant varieties of periodic points in all cases if there are invariants.

%%%%%%%%%%%%%%%%%%%%%%%%%%%%%%%%%%%%%%%%%%%%

\section{Derivation of recurrence equations}

If there is an invariant variety of periodic points of period $n$, every point on the variety can be an initial point of an $n$ periodic map. All images of the map are on the variety before the map returns to the initial point. Therefore this variety is clearly distinguished from the rest of $\hat{\mathbf C}^d$ and is reserved only for the maps of period $n$. In other words if the initial point is on this variety the map is always $n$ period.

This fact enables us to derive a recurrence equation once an invariant variety of periodic points is known. Let
\begin{equation}
x_j\ \rightarrow\ X_j=f_j(x_1,x_2,...,x_d),\qquad j=1,2,...,d
\label{X=f}
\end{equation}
be the map and (\ref{ivpp}) be the invariant variety of period $n$ of this map. We solve 
\begin{equation}
\gamma^{(n)}_\alpha(H_1(\mathbf{x}),H_2(\mathbf{x}),...,H_p(\mathbf{x}))=0,\qquad \alpha=1,2,...,l.
\label{gamma=0}
\end{equation}
for $l$ variables, say $x_{d-l+1},...,x_d$, and substitute them into $f_j,\ j=1,2,...,d-l$ of (\ref{X=f}). Then every initial point of the map $X_j=f_j,\ j=1,2,...,d-l$ is constrained on $v^{(n)}(\langle\gamma\rangle)$, thus we obtain a recurrence equation of period $n$. 

The simplest method to achieve this program is to find the $l$th elimination ideal of the functions $\{X_j-f_j,\ j=1,2,...,d-l\}$ and $\{\gamma^{(n)}_\alpha,\ \alpha=1,2,...,l\}$. If the ideal is generated by the functions $F_j^{(n)}$'s satisfying
\begin{equation}
F_j^{(n)}(X_1,X_2,...,X_{d-l},x_1,x_2,...,x_{d-l})=0,\qquad j=1,2,...,d-l,
\label{F=0}
\end{equation}
the recurrence equations are obtained by solving (\ref{F=0}) for $X_1,X_2,...,X_{d-l}$. Generally the solutions are not rational. But it does not cause any trouble, since the integrability of the map has been guaranteed from the begining. For all initial values $x_1,x_2,...,x_{d-l}$, the solutions of (\ref{F=0}) are periodic, hence are integrable.

For an illustration let us consider the map
\begin{equation}
(x,y)\rightarrow (X,Y)=\left(xy,\ {y(1+x)\over 1+xy}\right).
\label{(xy, y(1+x)/(1+xy))}
\end{equation}
This map has one invariant $H(x,y)=y(1+x)$ and the invariant variety of period 3 is given by the zeros of
\begin{eqnarray}
\gamma^{(3)}(x,y)&=&H^2+H+1\nonumber\\
&=&x^2y^2+2xy^2+y^2+xy+y+1.
\label{gamma of X=xy}
\end{eqnarray}
The 1st elimination ideal of the functions $X-xy$ and (\ref{gamma of X=xy}) is generated by the function
$$
F^{(3)}(X,x)=(x+1)^2X^2+x(x+1)X+x^2,
$$
from which we obtain two maps:
$$
x\rightarrow X=\left\{\begin{array}{l}
\displaystyle{\omega{x\over x+1}, }\cr
\qquad\qquad\qquad (\omega^3=1).\cr
\displaystyle{\omega^2{x\over x+1}}\cr
\end{array}\right.
$$
The iteration of the first map yields
$$
x\  \rightarrow\  \omega{x\over x+1}\ \rightarrow\  {\omega^2x\over -\omega^2 x+1}\ \rightarrow \ x,
$$
while the second map yields
$$
x\ \rightarrow\ \omega^2{x\over x+1}\ \rightarrow \ {\omega x\over -\omega x+1}\ \rightarrow\ x.
$$

In the rest of this section we would like to present various type of recurrence equations associated with invariant varieties of some integrable maps.\\

%%%%%%%%%%%%%%%%%%%%%%%%%%%%%%%%%%%

The $d$ dimensional Lotka-Volterra map is obtained by solving \cite{HTI}
\begin{equation}
X_j(1-X_{j-1})=x_j(1-x_{j+1}),\qquad j=1,2,...,d
\label{LV eq}
\end{equation}
for $\mathbf{X}=(X_1,X_2,...,X_d)$ under the conditions $x_{j+d}=x_j\ (j=1,2,...,d)$. The invariants of this map are given by
\begin{equation}
\left\{\begin{array}{cl}
H_k&={\sum}'_{j_1,j_2,...,j_k}x_{j_1}x_{j_2}\cdots x_{j_k}(1-x_{j_1-1})(1-x_{j_2-1})\cdots (1-x_{j_k-1})\cr
&\qquad\qquad\qquad\qquad ( k=1,2,...,[d/2] )\cr
r&=x_1x_2\cdots x_d\cr
\end{array}\right.
\label{H_k}
\end{equation}
Here the prime in the summation $\sum'$ of (\ref{H_k}) means that the summation must be taken over all possible combinations $j_1,j_2,...,j_k$ but excluding direct neighbours. The total number of the invariants is $p=[d/2]+1$, where $[d/2]=d/2$ if $d$ is even and $[d/2]=(d-1)/2$ if $d$ is odd. 

The invariant varieties have been derived in the cases of $d=3,4$ and 5 for some periods, explicitly\cite{SS}. In all examples the dimension of the invariant varieties is $p$. Hence the dimension of the recurrence equations is also $p$.

The 3 dimensional Lotka-Volterra map is given by, writing $(x_1,x_2,x_3)=(x,y,z)$,
\begin{equation}
 X=x{1-y+yz\over 1-z+zx},\quad Y=y{1-z+zx\over 1-x+xy},\quad Z=z{1-x+xy\over 1-y+yz},
\label{3LV}
\end{equation}
after solving (\ref{LV eq}) for $X,Y,Z$. 
There are two invariants
\begin{equation}
r=xyz,\qquad s=(1-x)(1-y)(1-z).
\label{r,s}
\end{equation}
The invariant varieties of periodic points have dimension 2 and are generated by the functions:
\begin{eqnarray}
\gamma^{(2)}&=&s+1\nonumber\\
%\label{period 2}\\
\gamma^{(3)}&=&r^2+s^2-rs+r+s+1\nonumber\\
%\label{period 3}\\
\gamma^{(4)}&=&r^3s+s^3-3rs^2+6r^2s+3rs-r^3+s\nonumber\\
\gamma^{(5)}&=& r^3s^4-r^3s^2-6r^4s^5+10r^3s^6+3s^5r+s^6+s^5+3r^4s^4-3r^5s^3\nonumber\\
&&
-6r^4s^3-r^6s^3+3r^5s^4+s^4+21s^4r^2+6s^4r+r^3s^7+s^7\nonumber\\
&&
+27s^5r^2-3s^6r-r^3s^5+21r^2s^6-10r^3s^3-6rs^7+s^8\nonumber\\&\vdots&
\label{3dLV gamma}
\end{eqnarray}
for the period 2,3,4,5,..., respectively. 

From these data we can derive a set of recurrence equations for each period. For the period 2 case we find
$$
F^{(2)}_1=(x-1)X-x,\qquad F^{(2)}_2=(y-1)Y-y,
$$
and the map is simply given as
$$
(x,y)\ \rightarrow\ \left({x\over x-1},\ {y\over y-1}\right)\ \rightarrow\ (x,y).
$$
In the period 3 case we obtain
\begin{eqnarray*}
F^{(3)}_1&=&(x^2-x+1)X^2+x(xy-2x+y+1)X+x^2(y^2-y+1)\\
F^{(3)}_2&=&\Big((3x^2-3x+1)y^2-(3x^2-5x+2)y+(x-1)^2\Big)Y^2\\
&-&y\Big((3x^2-2x+1)y-(2x-1)(x-1)\Big)Y+y^2(x^2-x+1).
\end{eqnarray*}
Since the solutions of $(F^{(3)}_1=0,\ F^{(3)}_2=0)$ are two folds the map has two routes:
\begin{eqnarray*}
(x,y)& \rightarrow& \left(\omega{x(y+\omega^2)\over x+\omega^2},\ {(1-\omega^2)y(x+\omega^2)\over 3xy+(x+y-1)\omega^2}\right)\\
&&
\rightarrow\ \left({(1-\omega^2)x(y+\omega^2)\over 3xy+(x+y-1)\omega^2}, \ \omega{y(x+\omega^2)\over y+\omega^2}\right)\ \rightarrow\ (x,y)\\
(x,y)& \rightarrow& \left(\omega^2{x(y+\omega)\over x+\omega},\ {(1-\omega)y(x+\omega)\over 3xy+(x+y-1)\omega}\right)\\
&&
\rightarrow\ \left({(1-\omega)x(y+\omega)\over 3xy+(x+y-1)\omega}, \ \omega^2{y(x+\omega)\over y+\omega}\right)\ \rightarrow\ (x,y)
\end{eqnarray*}
Similarly we can derive recurrence equations for larger periods, but their complicated expressions are not worth to be presented here for our purpose of this paper.\\

%%%%%%%%%%%%%%%%%%%%%%%%%%%%%%
The 4d Lotka-Volterra map $(x,y,z,u)\ \rightarrow\ (X,Y,Z,U)$ has three invariants. The invariant variety is generated by the function
$$
\gamma^{(2)}=H_1-2=x+y+z+u-xy-yz-zu-ux-2
$$
in the period 2 case, from which we derive the recurrence equation:
\begin{eqnarray*}
F^{(2)}_1&=&(1-x-z)X+x,\\
F^{(2)}_2&=&Y-y(1-x-z),\\
F^{(2)}_3&=&(1-x-z)Z+z.
\end{eqnarray*}
This provides an example of three dimensional map of period 2:
$$
(x,y,z)\ \rightarrow\ \left({x\over x+z-1},\ y(1-x-z),\ {z\over x+z-1}\right)\ \rightarrow\ (x,y,z).
$$
\ \\

The $N$ point Toda map is known equivalent to the $d=2N$ dimensional Lotka-Volterra map \cite{HTI}. In the case $N=3$, the map $(x,y,z,u,v,w) \rightarrow (X,Y,Z,U,V,W)$ is defined by
$$
X = y{zu+zx+wu\over yw+yz+vw},\quad 
Y = z{xv+xy+uv\over zu+zx+wu},\quad 
Z = x{yw+yz+vw\over xv+xy+uv},
$$$$
U = u{yw+yz+vw\over zu+zx+wu},\quad 
V = v{zu+zx+wu\over xv+xy+uv},\quad
W = w{xv+xy+uv\over yw+yz+vw}.
$$
Since this map has four invariants,
\begin{eqnarray*}
t_1&=&x+y+z+u+v+w,\\
t_2&=&xy+yz+zx+uv+vw+wu+xv+yw+zu,\\
t_3&=&xyz,\\
t'_3&=&uvw.
\end{eqnarray*}
the recurrence equations are expected to be four dimensional. The invariant variety of period 3 is given by the intersection of the functions\cite{SS}
$$
\gamma_1^{(3)}= t_1,\qquad \gamma_2^{(3)}= t_2.
$$
From this data we derive a four dimensional recurrence equation:
\begin{eqnarray*}
F^{(3)}_1&=&(x+y+u)X+(u+v+y)y,\\
F^{(3)}_2&=&(u+v+y)Y-(u+v+y)^2-(x-v)u,\\
F^{(3)}_3&=&,(u+v+y)U+(x+y+u)u, \\
F^{(3)}_4&=&(v-x)V+(u+v+y)v,
\end{eqnarray*}
or writing the solution explicitly, we find the map
$$
\left(\begin{array}{c}x\cr y\cr z\cr u\cr\end{array}\right)\rightarrow
\left(\begin{array}{c}
\displaystyle{-y{u+v+y\over x+y+u}}\cr\cr
\displaystyle{\phi(x,y,u,v)}\cr\cr
\displaystyle{-u{x+y+u\over u+v+y}}\cr\cr
\displaystyle{v{u+v+y\over x-v}}\cr\end{array}\right)
\rightarrow
\left(\begin{array}{c}
\displaystyle{\phi(x,y,u,v){x+y+u\over x-v}}\cr\cr
\displaystyle{x{u+v+y\over x-v}}\cr\cr
\displaystyle{u{x-v\over u+v+y}}\cr\cr
\displaystyle{-v{x+y+u\over x-v}}\cr\end{array}\right)
\rightarrow \left(\begin{array}{c}x\cr y\cr z\cr u\cr\end{array}\right)
$$
where 
$$
\phi(x,y,u,v)={(u+v+y)^2+(x-v)u\over u+v+y}.
$$
\ \\

As the last example we consider the Euler top. Let $(x,y,z)$ be the three components of the angular velocity of the Euler top. Then the map $(x,y,z)\rightarrow (X,Y,Z)$ satisfying
\begin{equation}
X=\alpha(Yz+Zy),\quad Y=\beta(Zx+Xz),\quad Z=\gamma(Xy+Yx)
\end{equation}
defines the discrete analog of the Euler top \cite{BLS}, if the parameters $(\alpha,\beta,\gamma)$ are related to the three moments of inertia $I,J,K$ of the top by
$$
\alpha={J-K\over 2I},\quad \beta={K-I\over 2J},\quad \gamma={I-J\over 2K}.
$$

This map has two invariants \cite{HT,BLS,HK}
$$
H_1={Ix^2+Jy^2+Kz^2\over 1-\beta\gamma x^2},\quad
H_2={I^2x^2+J^2y^2+K^2z^2\over 1-\beta\gamma x^2}, 
$$
from which we have found an invariant variety of periodic points \cite{SS2}
$$
v^{(3)}=\left\{{\mathbf x}\ \left|\ 3+\gamma {KH_1-H_2\over IJ}-\beta {JH_1-H_2\over KI}-\left(\alpha{IH_1-H_2\over 2JK}\right)^2=0\right.\right\}
$$
which is generated by the function
$$
\gamma^{(3)}=
(1+\beta\gamma x^2+\gamma\alpha y^2+\alpha\beta z^2)^2-4\alpha\beta\gamma(\alpha y^2z^2+\beta z^2x^2+\gamma x^2y^2)-4
$$
in the period 3 case. The recurrence equation of period 3 is then obtained as follows:
\begin{eqnarray*}
 F^{(3)}_{1}&=&\beta\Big((1-\alpha\gamma y^2)(\alpha\gamma y^2+\beta\gamma x^2-2-2q)X -(1-\alpha\gamma y^2+q)x\Big)^2\\
 &&-\alpha y^2(1-\alpha\gamma y^2+q)^2(\alpha\gamma y^2+\beta\gamma x^2-1-2q)\\
 F^{(3)}_{2}&=&\alpha\Big((1-\beta\gamma x^2)(\alpha\gamma y^2+\beta\gamma x^2-2-2q)Y- (1-\beta\gamma x^2+q)y\Big)^2\\
 &&-\beta x^2(1-\beta\gamma x^2+q)^2(\alpha\gamma y^2+\beta\gamma x^2-1-2q)
\end{eqnarray*}
where
$$
q=\sqrt{(1-\alpha\gamma y^2)(1-\beta\gamma x^2)}.
$$
The map has two routes,
$$
(x,y)\rightarrow \left\{\begin{array}{ccc}
(X_+,\ Y_+)&\rightarrow&(X_-,\ Y_-)\cr
(X_-,\ Y_-)&\rightarrow&(X_+,\ Y_+)\cr
\end{array}\right\}
\rightarrow (x,y),
$$
corresponding to the zeros of $(F^{(3)}_{1},\ F^{(3)}_{2})$:
\begin{eqnarray*}
X_\pm&=&{1-\alpha\gamma y^2+q\over 1-\alpha\gamma y^2}\ {x\pm y\sqrt{\alpha\beta(\alpha\gamma y^2+\beta\gamma x^2-1-2q)}\over \alpha\gamma y^2+\beta\gamma x^2-2-2q},\\
Y_\pm&=&{1-\beta\gamma x^2+q\over 1-\beta\gamma x^2}\ {y\pm x\sqrt{\alpha\beta(\alpha\gamma y^2+\beta\gamma x^2-1-2q)}\over \alpha\gamma y^2+\beta\gamma x^2-2-2q}.
\end{eqnarray*}
The two routes correspond to the forward and the backward maps starting from the same initial point. This means that the discrete Euler top can not start its 3 period motion unless the direction of the motion is informed.
%%%%%%%%%%%%%%%%%%%%%%%%%%%%%%%%%%%%%%%%%%%%%%

\section{Series of recurrence equations}

Let $f_1({\mathbf x}),f_2({\mathbf x}),...,f_{d-p}({\mathbf x}),\ H_1({\mathbf x}),...,H_p({\mathbf x})$ be some functions of ${\mathbf x}=(x_1,x_2,...,x_d)$. If $g_j({\mathbf x}),\ j=d-p+1,...,d$ are the solutions of
$$
H_i(f_1,f_2,...,f_{d-p},g_{d-p+1},...,g_d)=H_i(x_1,x_2,...,x_d),\quad i=1,2,...,p,
$$
they define a map
$$
{\mathbf x}\ \rightarrow\ {\mathbf X}=\Big(f_1({\mathbf x}),f_2({\mathbf x}),...,f_{d-p}({\mathbf x}),g_{d-p+1}({\mathbf x}),...,g_d({\mathbf x})\Big),
$$
in which $H_1({\mathbf x}),H_1({\mathbf x}),...,H_p({\mathbf x})$ are invariant.

For example, if $h({\mathbf x})$ is a function of $H_1,H_2,...,H_{d-1}$, and $f_1$ is given by
$$
f_1({\mathbf x})=hx_1(1-x_1),
$$
we obtain higher dimensional nonintegrable maps which reduce to the logistic map upon the elimination of $x_2,x_3,...,x_d$ by using the invariants. This idea enables us to consider many higher dimensional maps all together, just by studying a simple lower dimensional one.

%%%%%%%%%%%%%%%%%%%%%%%%%%%%%%%%%%%%%%%%%%%%%
\subsection{M\"obius map series}

Following to the above prescription we can derive higher dimensional integrable maps which reduce to the M\"obius map, if $a, b, h$ are some functions of the invariants $H_1,H_2,...,H_{d-1}$ and we define $f_1$ by
\begin{equation}
f_1({\mathbf x})=h{x_1+a\over 1+bx_1}.
\label{Mobius}
\end{equation}
The iteration of this map does not change the form of the map but only changes the functions $a,b,h$. Since we have assumed that these functions are dependent on the invariants $H_i({\mathbf x})$ alone, the initial values $(a,b,h)$ remain constant through the iteration. If we write 
\begin{equation}
X^{(n)}=h^{(n)}{x+a^{(n)}\over 1+b^{(n)}x}
\label{mobius}
\end{equation}
after $n$ steps, the $(n+1)$th parameters are related to the $n$th ones by
$$
a^{(n+1)}={a+a^{(n)}h^{(n)}\over h^{(n)}+ab^{(n)}},\quad 
b^{(n+1)}={b^{(n)}+bh^{(n)}\over 1+bh^{(n)}a^{(n)}},\quad 
h^{(n+1)}=h{h^{(n)}+ab^{(n)}\over 1+bh^{(n)}a^{(n)}},
$$
from which we can determine all parameters iteratively as functions of the initial values $(a,b,h)$.

The periodicity conditions of period $n$ for the map (\ref{Mobius}) are now satisfied if the parameters $(a,b,h)$ satisfy
\begin{equation}
(a^{(n+1)}, b^{(n+1)}, h^{(n+1)})=(a,b,h).
\label{(a_n+1, b_n+1, h_n+1)=(a,b,h)}
\end{equation}
From the construction it is clear that the periodicity conditions do not fix the values of the variable $x$ but impose some constranits on the parameters, hence on the invariants.

Solving (\ref{(a_n+1, b_n+1, h_n+1)=(a,b,h)}) iteratively we find the invariant varieties of periodic points as follows \cite{SS}
\begin{eqnarray}
v^{(2)}&=&\{{\mathbf x}| 1+h=0\}\nonumber\\
v^{(3)}&=&\{{\mathbf x}| 1+h+h^2+abh=0\}\nonumber\\
v^{(4)}&=&\{{\mathbf x}| 1+h^2+2abh=0\}\nonumber\\
v^{(5)}&=&\{{\mathbf x}| 1+h+h^2+h^3+h^4+abh(3+(4+ab)h+3h^2)=0\}\nonumber\\
v^{(6)}&=&\{{\mathbf x}| 1-h+h^2+3abh=0\}\nonumber\\
&\vdots&
\label{gamma of moebius}
\end{eqnarray}

According to our argument in \S 1 we should have recurrence equations corresponding to the invariant varieties (\ref{gamma of moebius}), one for each period. To obtain recurrence equations we must specify the invariants of the map in higher dimensions. Although the dimension of the map could be chosen arbitrary, we consider here two dimensions for the sake of simplicity. The number of the invariants is one in this case. Let $H(x,y)$ be the invariant. A two dimensional map, which reduces to (\ref{Mobius}), will be obtained if we fix $f_1(x,y)$ and $H(x,y)$ as functions of $(x,y)$. For this purpose we further assume simply that $a,b$ are constants and the function $f_1$ and the invariant $H$ are given by
$$
f_1(x,y)=H(x,y){x+a\over 1+bx},\qquad H(x,y)=y(1+bx).
$$
Solving $H(f_1,g)=H(x,y)$ for $g(x,y)$ we find a map
\begin{equation}
(x,y)\rightarrow (X,Y)=\left((x+a)y,\ y{1+bx\over 1+by(x+a)}\right).
\label{moebius 2dim}
\end{equation}
This includes (\ref{(xy, y(1+x)/(1+xy))}) as a special case.

Since we have already the information (\ref{gamma of moebius}) of the invariant varieties it is not difficult to derive a series of recurrence equations associated with the two dimensional map (\ref{moebius 2dim}), one for each period, as follows:
\begin{eqnarray*}
 F^{(2)}&=&(1+bx)X+x+a,\\
 F^{(3)}&=&(1+bx)^2X^2+(1+ab)(1+bx)(x+a)X+(x+a)^2,\\
 F^{(4)}&=&(1+bx)^2X^2+2ab(1+bx)(x+a)X+(x+a)^2,\\
 F^{(5)}&=&(1+bx)^4X^4+(1+3ab)(1+bx)^3(x+a)X^3\\
 &+&(1+4ab+a^2b^2)(1+bx)^2(x+a)^2X^2+(1+3ab)(1+bx)(x+a)^3X+(x+a)^4,\\
 F^{(6)}&=&(1+bx)^2X^2-(1-3ab)(1+bx)(x+a)X+(x+a)^2,\\
 &\vdots&
\end{eqnarray*}

To convince ourselves let us see some of the maps explicitly. The map of period 2 is generated by $F^{(2)}$, from which we find
$$
x\ \rightarrow\  -\ {x+a\over 1+bx}\ \rightarrow\ x.
$$
We notice that the generating functions of period 3, 4 and 6 cases are similar. There are a pair of routes for each period. The map in the period 3 case, for example, is given by
$$x\ \rightarrow\ \left\{
\begin{array}{ccc}
\displaystyle{-\mu_+{x+a\over 1+bx}}&\ \rightarrow \ 
-\ \displaystyle{{x+a\mu_-\over \mu_-+bx}}\cr\cr
\displaystyle{-\mu_-{x+a\over 1+bx}}&\ \rightarrow \ 
-\ \displaystyle{{x+a\mu_+\over \mu_++bx}}\cr
\end{array}\ \right\}\ \rightarrow\ x,
$$
where
$$
\mu_\pm={1\over 2}\Big(1+ab\pm\sqrt{(3+ab)(ab-1)}\Big).
$$
%%%%%%%%%%%%%%%%%%%%%%%%%%%%%%%%%%%%%%%%%%%%%%%%%
\subsection{Biquadratic map}

By studying various higher dimensional integrable maps which reduce to a one dimensional map $x\rightarrow X$, we found, in \cite{SS}, that many of them reduce not to the M\"obius map but to the `biquadratic map' defined by the equation:
\begin{equation}
aX^2x^2+b(X+x)Xx+c(X-x)^2+dXx+e(X+x)+f=0.
\label{S(X,x)=0}
\end{equation}
Here 
\begin{equation}
\mathbf{q}=(a,b,c,d,e,f)\in \mathbf{C}^6
\label{q}
\end{equation}
are functions of $d-1$ invariants. The function $f_1({\mathbf x})$ is determined by solving (\ref{S(X,x)=0}) for $X$.

Because of the symmetry of the equation (\ref{S(X,x)=0}) under the exchange of $X$ and $x$, the iteration of the map leaves the form of the map and changes only the parameters ${\mathbf q}$, as it was shown in \cite{SS}. After repeating the iteration $n$ times, the $(n+1)$th parameters ${\mathbf q}^{(n+1)}$ are determined by ${\mathbf q}^{(n)}$ and the initial values ${\mathbf q}$. Since the parameters are functions of the invariants alone, the periodicity conditions ${\mathbf q}^{(n+1)}={\mathbf q}$ impose some constraints among the invariants different for each period. 

Solving the periodicity conditions iteratively we have found \cite{SS} a series of invariant varieties of periodic points, one for each period. If $v^{(n)}=\{{\mathbf x}|\ \gamma^{(n)}=0\}$ is the invariant variety of period $n$, the generating functions $\gamma^{(n)}$ are given by
\begin{eqnarray}
 \gamma^{(3)}(\mathbf{q})&=&af-be-3c^2+cd,\nonumber\\
 \gamma^{(4)}(\mathbf{q})&=&2acf-adf+b^2f+ae^2-2c^3+c^2d-2bce,
\nonumber\\
 \gamma^{(5)}(\mathbf{q})&=&
a^3f^3+\Big(-cf^2d+2cfe^2+fde^2-3ebf^2-e^4-c^2f^2\Big)a^2\nonumber\\
 &+&\Big(-13c^4f+18c^3fd+de^3b+2cf^2b^2+7dc^2e^2-ce^2d^2-2ce^3b\nonumber\\
 &+&2c^2feb-7fd^2c^2-14c^3e^2+cd^3f+fb^2e^2+f^2db^2-ebd^2f\Big)a\nonumber\\
 &-&
cd^2b^2f-b^3e^3-4c^3deb+cdb^2e^2+13ec^4b-f^2b^4+7fb^2c^2d\nonumber\\
 &+&
c^4d^2-5c^5d+5c^6-2fb^3ec-e^2c^2b^2+eb^3df-14fb^2c^3,
\label{gamma_3(p)}
\end{eqnarray}
and so on.\\

In order to derive recurrence equations, we must specify the higher dimensional maps. As we have shown in \cite{SS, SS2} the symmetric version of the QRT map \cite{QRT}, the 3d Lotka-Volterra map of (\ref{3LV}), the discrete Euler top, a special case of the $q$-Painlev\'e IV map belong to this categoly. For example the 3d LV map (\ref{3LV}) is equivalent to the biquadratic map if we choose
\begin{eqnarray*}
&a=r+1,\quad b=s-2r-1,\quad c=r-s,&\nonumber\\
&d=s^2+rs+5r-2s+1,\quad e=-r(s+1),\quad f=0,&
\end{eqnarray*}
from which we could derive the invariant varieties (\ref{3dLV gamma}).\\

%%%%%%%%%%%%%%%%%%%%%%%%%%%%%%%%%
\subsection{The recurrence equations derived from the QRT map}

In the rest of this section we want to derive recurrence equations generated from the QRT map. Let us consider the two dimensional map
\begin{equation}
 (x,y)\rightarrow (X,Y)=\left(y,\ {\eta'(y)\rho''(y)-\rho'(y)\eta''(y)-x\Big(\rho'(y)\xi''(y)-\xi'(y)\rho''(y)\Big)\over
\rho'(y)\xi''(y)-\xi'(y)\rho''(y)-x\Big(\xi'(y)\eta''(y)-\eta'(y)\xi''(y)\Big)}\right).
\label{QRT 2d map}
\end{equation}
Here
\begin{eqnarray*}
\xi'(x):=a'x^2+b'x+c',\qquad\quad&\quad& \xi''(x):=a''x^2+b''x+c'',\\
\eta'(x):=b'x^2+(d'-2c')x+e',&\quad& \eta''(x):=b''x^2+(d''-2c'')x+e'',\\
\rho'(x):=c'x^2+e'x+f',\qquad\quad&\quad& \rho''(x):=c''x^2+e''x+f'',
\end{eqnarray*}
and  $\mathbf{q}'=(a',b',c',d',e',f')$ and $\mathbf{q}''=(a'',b'',c'',d'',e'',f'')$ are constants. If we write $(Y,y,x)$ as $(x^{(n+1)},x^{(n)},x^{(n-1)})$, this is nothing but the symmetric case of the well known QRT equation \cite{QRT}
$$
 x^{(n+1)}={\eta'(x^{(n)})\rho''(x^{(n)})-\rho'(x^{(n)})\eta''(x^{(n)})-x^{(n-1)}\Big(\rho'(x^{(n)})\xi''(x^{(n)})-\xi'(x^{(n)})\rho''(x^{(n)})\Big)\over
\rho'(x^{(n)})\xi''(x^{(n)})-\xi'(x^{(n)})\rho''(x^{(n)})-x^{(n-1)}\Big(\xi'(x^{(n)})\eta''(x^{(n)})-\eta'(x^{(n)})\xi''(x^{(n)})\Big)}.
%\label{QRT equation}
$$
The map (\ref{QRT 2d map}) has an invariant \cite{QRT}
\begin{equation}
H(x,y)=-\ {\xi'(x)y^2+\eta'(x)y+\rho'(x)\over \xi''(x)y^2+\eta''(x)y+\rho''(x)},
\label{h QRT}
\end{equation}
hence it can be reduced to one dimensional map $x\rightarrow X$. The calculation of the 2nd elimination ideal is rather trivial in this case. If $y$ and $Y$ are eliminated by using the invariant $H(x,y)=h$, the result we obtain is
\begin{equation}
\xi(x)X^2+\eta(x)X+\rho(x)=0
\label{phi(x)X^2+eta(x)X+rho(x)=0}
\end{equation}
where
$$
\xi(x):=ax^2+bx+c,\quad \eta(x):=bx^2+(d-2c)x+e,\quad
\rho(x):=cx^2+ex+f,
$$
with 
$$
\mathbf{q}=\mathbf{q}'+h\mathbf{q}''.
$$
If we identify $\mathbf{q}=(a,b,c,d,e,f)$ with those of (\ref{q}), the map (\ref{phi(x)X^2+eta(x)X+rho(x)=0}) is exactly the biquadratic map (\ref{S(X,x)=0}). \\

In the theory of the QRT map the formula (\ref{phi(x)X^2+eta(x)X+rho(x)=0}) is called an invariant curve\cite{QRT}. The problem of solving the equation (\ref{QRT 2d map}) is now converted to finding the coefficients of (\ref{phi(x)X^2+eta(x)X+rho(x)=0}) iteratively. Our general formula (\ref{gamma_3(p)}) gives us the explicit expressions of the invariant varieties of the symmetric QRT map (\ref{QRT 2d map}). Under these circumstances the problem of deriving the recurrence equations is quite simple. Namely we are already given in (\ref{gamma_3(p)}) the generating functions of invariant varieties in the form
\begin{equation}
\gamma^{(n)}\Big({\mathbf q}'+H(x,y){\mathbf q}''\Big)=0,\qquad n=3,4,5,....
\label{QRT gamma}
\end{equation}
The elimination of $y$ from the map (\ref{QRT 2d map}) and (\ref{QRT gamma}) amounts to replace $y$ by $X$ in  (\ref{QRT gamma}), since $X=y$. Thus we have found that the recurrence equations derived from the QRT map are generated by the functions
\begin{equation}
\gamma^{(n)}\Big({\mathbf q}'+H(x,X){\mathbf q}''\Big)=0,\qquad n=3,4,5,....
\label{QRT recurrence equations}
\end{equation}
The recurrence equation of period 3 is, for example,
\begin{eqnarray*}
 F^{(3)}&=&(a'+H(x,X)a'')(f'+H(x,X)f'')-(b'+H(x,X)b'')(e'+H(x,X)e'')\\
&&\qquad-3(c'+H(x,X)c'')^2+(c'+H(x,X)c'')(d'+H(x,X)d'').
\end{eqnarray*}

For some particular choices of the QRT parameters $({\mathbf q}', {\mathbf q}'')$, recurrence equations had been derived by the authors of \cite{HY}. Our formula (\ref{QRT recurrence equations}) provides such equations for all QRT parameters and for all periods.\\

%%%%%%%%%%%%%%%%%%%%%%%%%%%%%%%%%%%%%
\noindent
{\bf Acknowledgements}

The authors would like to thank the organizers of the SIDE VII meeting held in Melbourne, who extended us a kind hospitality during the meeting and gave an oppotunity to write this contribution.

%%%%%%%%%%%%%%%%%%%%%%%%%%%%%%%%%%%%%%%%
%%%%%%%%%%%%%%%%%%%%%%%%%%%%%%%%%%%%%

\end{document}